\documentclass[reqno,12pt]{amsart}
\usepackage{amssymb, psfig, epsf}
\numberwithin{equation}{section}
\def\be{\begin{equation}}
\def\ba{\begin{align}}
\def\bm{\begin{multline}}
\def\bfig{\begin{figure}[htb]}
\def\efig{\end{figure}}

\newcommand{\paper}[1]{{\it #1}, }
\newcommand{\journal}[4]{#1 {\bf #2}, #3 (#4)}

\newcommand{\HPA}{Helv. Phys. Acta}
\newcommand{\JPA}{J. Phys. A}
\newcommand{\JSP}{J. Stat. Phys.}

\newcommand{\PRB}{Phys. Rev. B}
\newcommand{\PRL}{Phys. Rev. Lett.}

\DeclareMathOperator{\Tr}{Tr}
\newtheorem{theorem}{Theorem}

\renewcommand{\leq}{\;\leqslant\;}
\renewcommand{\geq}{\;\geqslant\;}
\newcommand{\isdefby}{\;\doteqdot\;}

\newcommand{\dd}{{\rm d}}
\newcommand{\e}[1]{\,{\rm e}^{#1}\,}
\newcommand{\ii}{{\rm i}}
\newcommand{\sumtwo}[2]{\sum_{\substack{#1 \\ #2}}}

\newcommand{\abs}[1]{\lvert #1 \rvert}

\newcommand{\braket}[2]{\langle#1 | #2 \rangle}

\newcommand{\expval}[1]{\langle #1 \rangle_{\Lambda}}

\newcommand{\duha}[2]{( #1 , #2 )_{\Lambda}}
\newcommand{\ham}{H_{\Lambda}}
\newcommand{\vol}{\lvert \Lambda \rvert}

\begin{document}

\begin{quote}
\raggedleft
\end{quote}
\vspace{2mm}

\title{Charge density wave and quantum fluctuations in a molecular crystal}

\author[N. Macris and C.-A. Piguet]{N. Macris and C.-A. Piguet}

\maketitle

\begin{centering}
{\it Institut de Physique Th\'eorique \\
Ecole Polytechnique F\'ed\'erale de Lausanne \\
CH-1015 Lausanne, Switzerland\\}
\end{centering}

\thispagestyle{empty}

\vspace{10mm}

\noindent{\bf Abstract}
\vspace{5mm}

\noindent We consider an electron-phonon system in two and three dimensions on square, hexagonal and cubic lattices. The model is a modification of the standard Holstein model where the optical branch is appropriately curved in order to have a reflection positive Hamiltonian. Using infrared bounds together with a recent result on the coexistence of long-range order for electron and phonon fields, we prove that, at sufficiently low temperatures and sufficiently strong electron-phonon coupling, there is a Peierls instability towards a period two charge-density wave at half-filling. Our results take into account the quantum fluctuations of the elastic field in a rigorous way and are therefore independent of any adiabatic approximation. The strong coupling and low temperature regime found here is independent of the strength of the quantum fluctuations of the elastic field. 

\vspace{2mm}
\vspace{4mm}

\noindent Pacs: 63.20+Kr,63.70+h,67.40.Db
\newpage

\setcounter{page}{1}

\section{Introduction}

The Holstein model was originally introduced \cite{holstein59} to describe the motion of polarons in molecular crystals. Despite the simplifications introduced, it contains the essential features necessary to describe the interaction between lattice vibrations and itinerant electrons in a crystal. In the simplest version, the elastic field is modeled by a collection of Einstein oscillators with a single frequency associated to each site of the lattice and the electrons are treated as noninteracting fermions. This is a situation where itinerant electrons interact with a flat optical phonon branch. More sophisticated versions of molecular crystal models take into account a dispersion in the optical branch, as well as the Coulomb interaction between electrons. When a single electron is present (or when the electronic density is low), the Holstein model has been widely used to study the composite entity formed by the motion of the electron and its associated lattice deformation, the so-called "Holstein polaron". We refer to \cite{lowen88,gl91,aar92} for recent works on the subject.

Another situation of interest is the one where the electron density is large and where instabilities such as superconductivity (SC), charge density wave (CDW), spin density wave (SDW) occur. In the present work we focus on CDW formation for the Holstein model at half-filling. The prediction that such an instability may be present in electron-phonon systems goes back to Peierls \cite{peierls55} and Fr\"ohlich \cite{froehlich54}. They showed within the adiabatic, or more accurately static, approximation for the elastic field that, for one dimensional electron-phonon systems, a periodic modulation of the lattice commensurate to the electron filling would open a gap at the Fermi level. Remarkably, the electronic energy lowering caused by the gap opening is larger than the elastic energy cost and therefore a "Peierls instability" occurs. This phenomenon plays an important role in organic compounds such as polymer chains \cite{bcm92} and is for instance responsible for bond alternation in polyacetylene chains (CH)$_x$ \cite{ssh80} or annulenes. In two or three dimensional systems, this instability is not generic because of the more complicated shape of the Fermi surface. However, if the Fermi surface has nesting properties (for example on a square lattice at half-filling it is perfectly nested), the Peierls instability will occur for the same reasons than in one dimension. 

The analysis of Peierls and Fr\"ohlich and most of the subsequent work has been based on the adiabatic approximation for the elastic field. In many respects, it is still an open problem to determine when the Peierls instability survives against the quantum fluctuations of the elastic field. According to the analysis of Hirsch and Fradkin \cite{hf83} and to perturbative renormalization group calculations \cite{bc84}, the Peierls instability should not occur for weak electron-phonon coupling if the itinerant fermions are spinless but should occur for any coupling when spin is taken into account (at least at half-filling). However, recently this view has been challenged by numerical work \cite{jzw99} which predicts that the instability is destroyed at weak electron-phonon coupling even when spin is involved. For strong electron-phonon coupling, it is likely that the CDW forms in all cases.    

There exist a few rigorous studies of the question starting with a result of Kennedy and Lieb \cite{kl87} for the (static) Su-Schrieffer-Heeger (SSH) model. Their analysis shows rigorously that a period two ground state forms at half-filling consisting of alternating short and long bonds when electron-electron interaction is not taken into account. This result has been extended when a Hubbard interaction is added to the SSH Hamiltonian, by using a Jordan-Wigner mapping of the problem on a spin model \cite{ln94}. It is shown that at half-filling the ground state is either translation invariant or period two. The static Holstein model has also been the object of rigorous work in one, two and three dimensions. It is known that for the half-filled situation the ground state is period two in any dimension and that this symmetry breaking persists for low temperatures for dimensions greater or equal to two \cite{lm94b}. Away from half-filling, it has been proved in the one dimensional static Holstein model that the Peierls instability also occurs for rational densities \cite{bgm98}. When quantum fluctuations of the elastic field are taken into account Benfatto, Gallavotti and Lebowitz \cite{bgl95} have shown, using a rigorous version of the renormalization group analysis valid for spinless electrons in one dimension, that there is no instability for weak electron-phonon coupling and the electron correlation functions are those of a Luttinger liquid. We note that their analysis holds for any electron filling but breaks down when electron spin is included. Also it does not cover the case of large electron-phonon coupling. A very general result of Freericks and Lieb \cite{fl} asserts that for any even number of electrons on a finite lattice, the ground state of the Holstein model is unique and has zero total spin. This property is compatible with the presence or absence of a Peierls or other instabilities and with the uniqueness or non-uniqueness of ground or Gibbs states in thermodynamic limit, as long as the system is in a singlet state. 
 
In this work, we consider the two and three dimensional situation for a modified Holstein model taking into account quantum fluctuations of the elastic field exactly. Our model consists of non-interacting spinless electrons on square, hexagonal or cubic lattices interacting with an optical phonon branch which is appropriately curved. For this special phonon branch, we are able to use infrared bounds together with a recent result on the coexistence of  long-range order for electron and phonon fields, to prove that a period two CDW forms at strong coupling and half-filling. Our result shows that at least for some regime of parameters the Peierls instability survives quantum and thermal fluctuations. The regime that we identify is independent of the strength of the quantum fluctuations. Our use of infrared bounds is based on the recent observation that some fermion Hamiltonians have the reflection positivity property if the hopping matrix satisfies some sign conditions \cite{lieb94,mn96}. These conditions are most easily expressed in terms of the "magnetic flux" through a plaquette, i.e. the phase of the product of hopping amplitudes around an elementary loop of the lattice. For a square or cubic lattice it turns out that the magnetic flux through squares must be equal to $\pi$, whereas for hexagonal lattice the magnetic flux through hexagons must vanish. We note that it is not necessary to interpret this sign condition in terms of magnetic flux. For example in some situations the arrangement of orbitals of neighboring atoms may lead to overlap integrals having these sign properties. Such a situation has been encountered in coupled polymer chains \cite{bm88} and has also been the subject of speculations in the chemistry literature \cite{dd75}.

The paper is organized as follows. In section 2, we give the formal definition of the model, the main results and we outline the general strategy of the proof. Section 3 deals with the coexistence of long-range order for electron and phonon fields in a form which relates their associated correlation functions. This is then combined with the infrared bounds derived in section 4 to complete the proof for the square and cubic lattices in section 5. The necessary modifications for the hexagonal lattice are given in section 6. Finally we conclude by a discussion of a few open problems, that may be studied within the present formalism.

\section{Definition of the model and main results}

We consider a system of electrons and phonons interacting on a finite hypercubic 
lattice $N\times\ldots\times N$ with $N$ even, in $d$ dimensions and with 
periodic boundary conditions. We call $\vol=N^d$ the 
volume of the system. The Hamiltonian under consideration has the form
\be
\ham=\ham^e+\ham^{ph}+\ham^{int}.
\label{ham}
\end{equation}
$\ham^e$ is the kinetic Hamiltonian of spinless electrons 
\be
\ham^e=\sum_{x,y\in\Lambda}t_{xy}c_{x}^{+}c_{y}
\label{elec}
\end{equation}
where $c_{x}^{+}$ and $c_x$ are the
creation and annihilation operators for an electron at site $x$. They satisfy 
the canonical anticommutation relations $\{c_{x}^{+},c_{y}^{+}\}=\{c_{x},c_{y}\}=0$ and
$\{c_{x}^{+},c_{y}\}=\delta_{xy}$. The hopping amplitudes $t_{xy}=0$ for $\abs{x-y}\neq 1$ and  
$t_{xy}=\e{\ii \theta_{xy}}$ for $\abs{x-y}= 1$ where the phases
$\theta_{xy}$ are chosen in such a way 
that there is a magnetic flux $\pi$ across each elementary plaquette of the
lattice, i.e. $\sum_{{\rm plaquette}}\theta_{xy}=\pi$. We choose $N$ to be
even and, as explained in the introduction, impose such an external magnetic flux in order to have a Hamiltonian
which exhibits the reflection positivity property.

The Hamiltonian for the phonons $\ham^{ph}$ is 
\be
\ham^{ph}=\sum_{k}\omega(k)(b_{k}^{+}b_{k}+1/2)
\label{phon}
\end{equation}
where the creation and annihilation operators for a phonon with wave vector $k=(k_1,\ldots,k_d)$ satisfy commutation relations $[b_{k}^{+},b_{k'}^{+}]=[b_{k},b_{k'}]=0$. All $k$-sums are over the first Brillouin zone associated with the lattice. In the standard Holstein model one has a flat dispersion relation $\omega(k)=\omega=\sqrt{\kappa/m}$ where $m$ is a mass and $\kappa$ the oscillator constant \cite{holstein59}. Here we consider a slightly curved dispersion relation of the form
\be
\omega(k)=\sqrt{\omega^2+\frac{8\alpha}{m}(d+\sum_{i=1}^d\cos{k_i})}
\label{omega}
\end{equation}
where $\alpha$ is some positive parameter, the reason being that (\ref{omega}) is compatible with reflection positivity. In fact there is a general class of dispersion relations that we could allow, the simplest of which is (\ref{omega}), but we will not pursue this issue here. 

For the electron-phonon interaction we choose the simple form 
\be
\ham^{int}=g\sum_{k}\frac{1}{\sqrt{2m\omega(k)}}(b_{-k}+b_{k}^+)\rho_k
\label{inte}
\end{equation}
where $\rho_k$ is the Fourier transform of the electronic density $n_x-1/2=c_x^+c_x-1/2$
\be
\rho_k=\sum_{x\in\Lambda}\frac{\e{-\ii k\cdot x}}{\sqrt{\vol}}(n_x-1/2).
\label{fourier}
\end{equation}

It turns out to be more convenient to use the direct space language. We introduce the conjugate variables $s_x$ and $p_x$ with $[s_x,p_y]=\ii\delta_{xy}$,
\begin{align}
s_x&=\sum_{k}\frac{\e{\ii k\cdot x}}{\sqrt{\vol}}\frac{1}{\sqrt{2m\omega(k)}}
(b_k+b_{-k}^{+})\nonumber\\
p_x&=\sum_{k}\frac{\e{\ii k\cdot x}}{\sqrt{\vol}}\frac{1}{2\ii}\sqrt{2m\omega(k)}
(b_k-b_{-k}^{+}).
\end{align}
In these variables, the Hamiltonian for the phonons has the form
\be
\ham^{ph}=\sum_{x\in\Lambda}\frac{p_{x}^{2}}{2m}
+\frac{\kappa}{2}\sum_{x\in\Lambda}s_{x}^{2}+\alpha\sum_{\abs{x-y}=1}
(s_{x}+s_{y})^{2}
\end{equation}
and the electron-phonon interaction is on-site
\be
\ham^{int}=g\sum_{x\in\Lambda}s_{x}(n_{x}-\frac{1}{2}).
\end{equation}
The Hamiltonian (\ref{ham}) is invariant under the following unitary transformation $s_x\rightarrow -s_x$, $c_x^+\rightarrow (-1)^{\abs{x}}c_x$, $c_x\rightarrow (-1)^{\abs{x}}c_x^+$. 
As a consequence, we have that the thermal expectation value (denoted by $\expval{-}$) of the number of electrons is half the number of sites and the one of the $s_x$ variables is zero: $\expval{n_x}=\frac{1}{2}$, $\expval{s_x}=0$ for any temperature $\beta^{-1}$.

In this work, we are interested in the two long-range orders for the $s_x$ and $n_x$ variables at $k_0=(\pi,\pi,\pi)$ in three dimensions and $k_0=(\pi,\pi)$ in two dimensions, i.e. we will show that there exists $\epsilon_1$ and $\epsilon_2$ independent of $\Lambda$ such that
\be
\frac{1}{\vol}\expval{s_{-k_0}s_{k_0}}=\frac{1}{\vol^2}\sum_{x,y\in\Lambda}(-1)^{\abs{x}+\abs{y}}\expval{s_xs_y}\geq\epsilon_1>0
\label{first}
\end{equation}
and
\be
\frac{1}{\vol}\expval{\rho_{-k_0}\rho_{k_0}}=
\frac{1}{\vol^2}\sum_{x,y\in\Lambda}(-1)^{\abs{x}+\abs{y}}\expval{(n_x-1/2)(n_y-1/2)}\geq\epsilon_2>0
\label{second}
\end{equation}
for sufficiently large $\Lambda$ and some range of parameters. More precisely we show that long-range order is present on the square lattice at zero temperature if $(\kappa+32\alpha)/g\sqrt{\alpha}\leq 0.19$ and at sufficiently low temperatures if  $(\kappa+48\alpha)/g\sqrt{\alpha}\leq 0.22$ for a cubic lattice. It is of interest to notice that the strong coupling regime obtained here is independent of the mass $m$ of the oscillators. 

On the hexagonal lattice (section 6) we obtain a similar result for zero magnetic flux (i.e. all hopping terms have the same sign), namely (\ref{first}) and (\ref{second}) are valid at zero temperature for $(\kappa+24\alpha)/g\sqrt{\alpha}\leq 0.18$.

It has to be noticed that in the two dimensional cases (square and hexagonal lattices) our result is limited to zero temperature but this is probably an artifact of our technique. Since the symmetry breaking involved in this model is discrete (Ising like), we expect the result to be valid also at low temperatures.

One would think that the adjunction of the $\alpha$-term in the phonon Hamiltonian is able to create the long-range order (\ref{first}) even at $g=0$ since it has tendency to force two nearest neighbors $s_x$ and $s_y$ to have opposite signs or, in the phonon language, the mode $k_0$ is the most favorable energetically since $\omega(k)$ is minimum for $k=k_0$. This is in fact not the case. Indeed we can compute explicitly in the case $g=0$ the expectation value we are interested in. We find
\be
\frac{1}{\vol}\expval{s_{-k_0}s_{k_0}}=\frac{1}{\vol^2}\sum_{x,y\in\Lambda}(-1)^{\abs{x}+\abs{y}}\expval{s_xs_y}=
\frac{1}{\vol}\frac{1}{2m\omega}\left(1+\frac{2}
{\e{\beta\omega}-1}\right)
\label{hol3}
\end{equation}
showing that there is no long-range order since (\ref{hol3}) tends to zero as $\vol\rightarrow\infty$. We can thus conclude that the long-range orders will appear only if the electron-phonon interaction is present. For this reason, this modified model provides an example of long-range order induced by the electron-phonon interaction in the presence of realistic quantum fluctuations. 

Let us summarize the main steps for the strategy of the proof. First, in section 3, we establish the coexistence of the two long-range orders (\ref{first}) and (\ref{second}). For this we use results developed in a previous paper \cite{mp99} with which we can get a relation between the two-point Duhamel functions $\duha{s_{-k}}{s_k}$ and $\duha{\rho_{-k}}{\rho_k}$ for any $k$. If we use this relation at $k=k_0$, this shows the coexistence we are looking for. However, in the next steps of the proof, we will also need this relation for $k\neq k_0$. In section 4, we find an upper infrared bound for the Duhamel two-point functions $\duha{s_{-k}}{s_k}$ using reflection positivity and, with the help of the previously mentioned relation, we deduce an infrared bound for $\duha{\rho_{-k}}{\rho_k}$. In section 5, we use a theorem of Dyson-Lieb-Simon \cite{dls78} to obtain an upper bound on the two-point correlation function $\expval{\rho_{-k}\rho_k}$ and then the sum rule
\be
\sum_{k}\expval{\rho_{-k}\rho_k}=\sum_{x\in\Lambda}\expval{(n_x-1/2)^2}=\frac{\vol}{4}
\label{sumru}
\end{equation}
to prove the existence of long-range order (\ref{second}) for the electronic variables at $k_0$. A more direct way would have been to combine the infrared bound on $\duha{s_{-k}}{s_k}$ with a sum rule for the $s_x$ variables:
\be
\sum_{k}\expval{s_{-k}s_k}=\sum_{x\in\Lambda}\expval{s_x^2}.
\label{sumru3}
\end{equation}
It would be sufficient to have a lower bound for this quantity, but we were unable to find a convenient one. We will come back to this point at the end of section 5.

\section{Coexistence of long-range orders for electron and phonon fields}

In a previous paper, we have developed a general formalism to discuss the possible coexistence of two long-range orders in a quantum system at finite temperature \cite{mp99} which we summarize below. 
The Duhamel two-point function of two operators $F$ and $G$ is defined by
\be
(F,G)_{\Lambda}=\frac{1}{Z_{\Lambda}}\int_{0}^{1}\dd v
\Tr{\left(\e{-v\beta\ham}F\e{-(1-v)\beta\ham}G\right)}.
\end{equation}
The Duhamel two-point function $\duha{F^+}{F}$ is related to the symmetrized two-point correlation function through upper and lower bounds \cite{dls78}
\be
\frac{1}{2}\expval{F^{+}F+FF^{+}}f(h_{\Lambda}(F))\leq\duha{F^+}{F}\leq\frac{1}{2}\expval{F^{+}F+FF^{+}}
\label{upperlowerbound}
\end{equation}
where
\be
h_{\Lambda}(F)=
\frac{\expval{[F^{+},[\beta H_{\Lambda},F]]}}
{2\expval{F^{+}F+FF^{+}}}
\end{equation}
and the function $f(u)$ is defined implicitly for $u>0$ by
\be
f(u\tanh{u})=\frac{1}{u}\tanh{u}.
\end{equation}
The function $f(u)$ is continuous, convex and strictly decreasing with
$\lim_{u\rightarrow 0}f(u)=1$ and $\lim_{u\rightarrow\infty}f(u)=0$.

Let us suppose that the following commutation relation is satisfied between three local operators $A_x$, $B_x$ and $C_x$:
\be
[\ham,A_x]=\mu B_x+\nu C_x
\label{commutation}
\end{equation}
where $\mu$ and $\nu$ are two complex numbers. Let $A_k$, $B_k$, $C_k$ the Fourier transforms defined as in (\ref{fourier}). As a consequence of (\ref{commutation}) we have the following relation between the Duhamel two-point functions of $B_k$ and $C_k$ \cite{mp99}
\be
\left(\abs{\mu}\sqrt{
\duha{B_k^+}{B_k}}-\abs{\nu}\sqrt{\duha{C_k^+}{C_k}}\right)^2
\leq\frac{1}{\beta}\expval{[A_k^+,K_k]}
\label{coex1}
\end{equation}
where $K_k=[\ham,A_k]$.

With the help of inequalities (\ref{upperlowerbound}) and (\ref{coex1}), we deduce the following general theorem of coexistence \cite{mp99}:
\begin{theorem}
\label{th1}
Assume there exist three local observables $A_x$, $B_x$, $C_x$ satisfying
(\ref{commutation}). Suppose also that, for a given $k$, there exist 
three positive constants $a_{k}$, $b_{k}$ and $c_{k}$
independent of $\vol$  such that
$\expval{[A_k^+,[\ham,A_k]]}\leq a_k$,
$\expval{[B_k^+,[\ham,B_k]]}\leq b_k$ and 
$\expval{[C_k^+,[\ham,C_k]]}\leq c_k$.
Then for any temperature $\beta^{-1}$ and any sufficiently large $\Lambda$, there exists $\epsilon_1$ independent of $\Lambda$ such that
\be
\frac{1}{\vol}\expval{B_k^+B_k+B_kB_k^+}\geq\epsilon_1 > 0
\label{coex2}
\end{equation}
if and only if there exists $\epsilon_2$ independent of $\Lambda$ such that 
\be
\frac{1}{\vol}\expval{C_k^+C_k+C_kC_k^+}\geq\epsilon_2 > 0.
\label{coex3}
\end{equation}
\end{theorem}

To apply these results to our Hamiltonian (\ref{ham}), we note the following commutation relation 
\be
[H,p_x]=\ii \big[(\kappa+8\alpha
d)s_x+4\alpha\sumtwo{y}{\abs{y-x}=1}s_{y}\big]+\ii
g(n_x-1/2)=\ii S_x+\ii g(n_x-1/2).
\label{hol4}
\end{equation}
Since the Fourier transform of $S_x$ is 
\be
S_{k}=\big[\kappa+4\alpha(2d+2\sum_{i=1}^d\cos k_i)\big]s_k
\end{equation}
(\ref{coex1}) becomes
\be
\left(\sqrt{\duha{S_{-k}}{S_k}}-g\sqrt{\duha{\rho_{-k}}{\rho_k}}\right)^{2}\leq
\frac{1}{\beta}\expval{[p_{-k},[H,p_k]]}.
\label{hol5}
\end{equation}
The double commutator on the right-hand side is simply a constant equal to $\kappa+4\alpha(2d+2\sum_{i}\cos k_i)$. 
Finally, (\ref{hol5}) reads
\bm
\left([\kappa+4\alpha(2d+2\sum_{i=1}^d\cos k_i)]
\sqrt{\duha{s_{-k}}{s_k}}-g\sqrt{\duha{\rho_{-k}}{\rho_k}}\right)^{2}\\
\leq\frac{1}{\beta}[\kappa+4\alpha(2d+2\sum_{i=1}^d\cos k_i)].
\label{hol7}
\end{multline}
We will need inequality (\ref{hol7}) in the next sections for $k\neq k_0$. For the case
under consideration in this section, i.e. to prove the coexistence of the two
long-range orders at $k=k_0$, we evaluate (\ref{hol7}) at $k=k_0$ to get
\be
\kappa\sqrt{\duha{s_{-k_0}}{s_{k_0}}}\geq g\sqrt{\duha{\rho_{-k_0}}{\rho_{k_0}}}-
\sqrt{\frac{\kappa}{\beta}}.
\label{ineqs-npi}
\end{equation}
We can now get information about the two-points
correlation functions using the lower and upper bounds (\ref{upperlowerbound}). Since the $s_x$ are commuting variables
\be
\duha{s_{-k_0}}{s_{k_0}}\leq\expval{s_{-k_0}s_{k_0}}.
\label{deux}
\end{equation}
For the lower bound involving the electronic densities, we
need to compute the double commutator
\be
[n_{-k},[H,n_k]]=\frac{1}{\vol}\sum_{x,y\in\Lambda}t_{xy}(2\cos
[k\cdot(y-x)]-2)c_{x}^{+}c_{y}.
\label{dc}
\end{equation}
The expectation value of (\ref{dc}) is bounded by $8d$ 
since $\abs{t_{xy}(2\cos
[k(y-x)]-2)}\leq 4$ and $\abs{\expval{c_{x}^{+}c_{y}}}\leq 1$.
Since the function $f$ is decreasing, we can replace the double commutator by $8d$ and get 
\be
\duha{\rho_{-k_0}}{\rho_{k_0}}\geq
\expval{\rho_{-k_0}\rho_{k_0}}f\left(\frac{4\beta d}{\expval{\rho_{-k_0}\rho_{k_0}}}\right)
\label{trois}
\end{equation}
where we also used the fact that the $n_x$ variables commute.  
Finally, from (\ref{ineqs-npi}), (\ref{deux}) and (\ref{trois}) 
\be
\kappa\sqrt{\expval{s_{-k_0}s_{k_0}}}\geq
 g\sqrt{\expval{\rho_{-k_0}\rho_{k_0}}f\left(\frac{2\beta d}{\expval{\rho_{-k_0}\rho_{k_0}}}\right)}-
\sqrt{\frac{\kappa}{\beta}}.
\label{because}
\end{equation}
This last inequality shows that if
$\expval{\rho_{-k_0}\rho_{k_0}}=O(\vol)$, we have also that 
$\expval{s_{-k_0}s_{k_0}}=O(\vol)$ since the function $f(x)$ goes to 1 as $x$ goes to zero. In section 4, we
will prove the existence of the long-range order for the electronic densities at $k=k_0$.
Because of (\ref{because}), this implies the existence of the long-range order in the
phonon variables.

To close this section, let us notice that the commutation relation (\ref{hol4}) in the case
$g=0$ immediately leads to the absence of long-range order in the phonon
variables for any $k$. This is consistent with the exact computation
(\ref{hol3}).

\section{Infrared bounds}

The first step is to derive an infrared  bound for the Duhamel two-point function of 
the phonon variables $\duha{s_{-k}}{s_{k}}$ for any $k\neq{k_0}$ by using reflection positivity. We couple the phonon field in the Hamiltonian to a real symmetric field $h_{xy}$, $h_{xy}=h_{yx}$, as follows 
\bm
\ham(\{h_{xy}\})=\sum_{x\in\Lambda}\frac{p_{x}^{2}}{2m}
+\frac{\kappa}{2}\sum_{x\in\Lambda}s_{x}^{2}+\alpha\sum_{\abs{x-y}=1}
(s_{x}+s_{y}-h_{xy})^{2}\\
+\sum_{\abs{x-y}=1}t_{xy}c_{x}^{+}c_{y}+U\sum_{x\in\Lambda}s_{x}(n_{x}-1/2).
\label{hol10}
\end{multline}
In the form (\ref{hol10}) $\ham(\{h_{xy}\})$ is not reflection positive. However it can be transformed into a reflection positive form after a succession of appropriate transformations. We stress that these transformations cannot be done in general for fermions systems because the kinetic term usually does not have the correct sign properties, except if a flux $\pi$ is imposed through each plaquette of the square or cubic lattice. This has been shown in detail recently \cite{mn96} and using these results one deduces in a standard way that the partition function $Z_{\Lambda}$ associated to (\ref{hol10}) satisfies $Z_{\Lambda}(\{h_{xy}\})\leq Z_{\Lambda}$. 

We can now expand the left hand side of this last inequality to second order in $\{h_{xy}\}$ to obtain
\be
\duha{\sum_{\abs{x-y}=1}h_{xy}s_{x}}{\sum_{\abs{x'-y'}=1}h_{x'y'}s_{x'}}\leq
\frac{1}{8\beta\alpha}\sum_{\abs{x-y}=1}h_{xy}^{2}.
\label{ineqh}
\end{equation}
This inequality has been derived for real symmetric $h_{xy}$ but it is possible to extend it to complex numbers: 
\be
\duha{\sum_{\abs{x-y}=1}h_{xy}^{*}s_{x}}{\sum_{\abs{x'-y'}=1}h_{x'y'}s_{x'}}\leq
\frac{1}{8\beta\alpha}\sum_{\abs{x-y}=1}\abs{h_{xy}}^{2}.
\label{analog}
\end{equation}
For each $k$ in the first Brillouin zone, we choose $h_{xy}$ to be 
\be
h_{xy}=\frac{\e{-\ii k\cdot x}+\e{-\ii k\cdot y}}{\sqrt{\vol}}.
\end{equation}
This immediately leads to the infrared bound for the phonon variables
\be
\duha{s_{-k}}{s_{k}}\leq\frac{1}{4\beta\alpha}\frac{1}{(2d+2\sum_{i}\cos k_{i})}.
\label{duha-s}
\end{equation}

From the relation (\ref{hol7}) we transfer the information contained in (\ref{duha-s}) onto the Duhamel two-point function for the electronic densities
\bm
g\sqrt{\duha{\rho_{-k}}{\rho_k}}\leq [\kappa+4\alpha(2d+2\sum_{i}\cos k_i)]\sqrt{\duha{s_{-k}}{s_k}}\\
+\left(\frac{1}{\beta}[\kappa+4\alpha(2d+2\sum_{i}\cos k_i)]\right)^{1/2}
\end{multline}
and inserting (\ref{duha-s})
\ba
g&\sqrt{\duha{\rho_{-k}}{\rho_k}}\leq \frac{\kappa+4\alpha(2d+2\sum_{i}\cos k_i)}{\sqrt{4\beta\alpha(2d+2\sum_{i}\cos k_{i})}}+\left(\frac{\kappa+4\alpha(2d+2\sum_{i}\cos k_i)}{\beta}\right)^{1/2}\nonumber\\
&\leq \frac{\kappa+16\alpha d}{\sqrt{4\beta\alpha(2d+2\sum_{i}\cos k_{i})}}+\frac{\sqrt{\kappa+16\alpha d}}{\sqrt{\beta}}\nonumber\\
&\leq \frac{\kappa+16\alpha d+\sqrt{\kappa+16\alpha d}\sqrt{4\alpha(2d+2\sum_{i}\cos k_{i})}}
{\sqrt{4\beta\alpha(2d+2\sum_{i}\cos k_{i})}}
\leq \frac{2(\kappa+16\alpha d)}{\sqrt{4\beta\alpha(2d+2\sum_{i}\cos k_{i})}}.
\end{align}
In summary, we have also an infrared bound for the electron density 
\be
\duha{\rho_{-k}}{\rho_k}\leq \frac{(\kappa+16\alpha d)^2}{\beta\alpha g^2(2d+2\sum_{i}\cos k_{i})}\isdefby \frac{B_k}{\beta}.
\label{hol8}
\end{equation}

\section{Proof of long-range order for the square and cubic lattices}

Inverting the lower bound in (\ref{upperlowerbound}) we obtain an estimate for the two-point correlation function of the electronic variables 
\be
\expval{\rho_{-k}\rho_k}\leq
\frac{1}{2}\sqrt{B_kC_k}\coth\sqrt{\frac{\beta^2C_k}{4B_k}}
\label{bk}
\end{equation}
where $B_k$ is defined in (\ref{hol8}) and $C_k$ is the upper bound of the double commutator (\ref{dc})
\be
\expval{[n_{-k},[H,n_k]]}\leq 8d\isdefby C_k.
\end{equation}
Using (\ref{bk}) and the sum rule (\ref{sumru}) in the limit of large volumes, we get
\ba
\frac{1}{\vol}\expval{\rho_{-k_0}\rho_{k_0}}&\geq
\frac{1}{4}-\frac{1}{(2\pi)^d}\int\dd^d k \frac{\sqrt{2}(\kappa+16\alpha d)}{g}\sqrt{\frac{d}{\alpha(2d+2\sum_{i}\cos
k_i)}}\cdot\nonumber\\
&\coth\left[\frac{\sqrt{2}\beta g}{(\kappa+16\alpha d)}\sqrt{d\alpha(2d+2\sum_{i}\cos
k_i)}\right].
\label{integral}
\end{align}
Inequality (\ref{integral}) shows that there is long-range order whenever the right hand side is strictly positive. 
Because of theorem \ref{th1} the long-range order occurs both in $\frac{1}{\vol}\expval{\rho_{-k_0}\rho_{k_0}}$ and in $\frac{1}{\vol}\expval{s_{-k_0}s_{k_0}}$.

The integral in (\ref{integral}) can be computed numerically (when $\beta=\infty$) which leads to the condition stated in section 2 for three dimensions
\be
\frac{(\kappa+48\alpha)}{g\sqrt{\alpha}}\leq 0.22,\;\;\;d=3,\hspace{2mm}{\rm and}\hspace{2mm}{\rm large}\hspace{2mm}\beta.
\label{notm}
\end{equation}
In two dimensions, (\ref{integral}) is not a convergent integral. One must first take the limit $\beta\rightarrow\infty$ which yields an inequality similar to (\ref{integral}) but with the $\coth$ replaced by 1. Computing the remaining integral leads to the condition
\be
\frac{(\kappa+32\alpha)}{g\sqrt{\alpha}}\leq 0.19,\;\;\;d=2,\hspace{2mm}{\rm and}\hspace{2mm}\beta=\infty.
\end{equation}
If we fix $\alpha$ and $\kappa$, we see that there will be long-range order for large values of $g$. On the other hand, if we fix $g$ and $\alpha$, there will be long-range order for sufficiently small $\kappa$. These two results are intuitively plausible: there is long-range order for large electron-phonon interaction and also the oscillator constant should not be too large. Finally, if we fix $g$ and $\kappa$, the situation for $\alpha$ is more subtle. We see that it should not be too large and also it should not be too small. In particular, we are not allowed to set $\alpha=0$ and thus we cannot conclude anything concerning the Holstein model with a flat branch. 

A surprising point is the absence of the mass of the phonons in the derived conditions. It would not have been the case if we would have considered the sum rule (\ref{sumru3}). Indeed, in this case, we would have transferred inequality (\ref{duha-s}) onto the two-point correlation function for the $s_x$ variables. For this, we need to compute the double commutator $[s_{-k},[\ham,s_k]]=\frac{1}{m}$
that now depends on $m$. With this method, we obtain the inequality
\ba
\frac{1}{\vol}\expval{s_{-k_0}s_{k_0}}&\geq \expval{s_x^2}-\frac{1}{(2\pi)^d}\int\dd^d k \frac{1}{2}\sqrt{\frac{1}{4\alpha m(2d+2\sum_{i}\cos k_i)}}
\cdot\nonumber\\
&\coth\sqrt{\frac{\beta^2}{16m\alpha(2d+2\sum_{i}\cos k_i)}}.
\end{align}
Unfortunately we are unable to use it since we do not know how to derive a lower bound for $\expval{s_x^2}$. However, this discussion suggests that a lower bound on $\expval{s_x^2}$ would yield existence of long-range order for another region in the space of parameters which would now depend on $m$.

\section{The hexagonal lattice}

Using the same technique, we can prove a similar result for a two-dimensional hexagonal lattice at zero temperature. In this case however, we do not need to impose an external magnetic flux on the system. We can thus choose real positive hopping amplitudes  $t_{xy}=1$ for $\abs{x-y}=1$ and in direct space the Hamiltonian is given again by the sum of (\ref{elec}), (\ref{phon}), (\ref{inte}). 
This Hamiltonian is still invariant under a particle-hole transformation and we have $\expval{n_x}=\frac{1}{2}$, $\expval{s_x}=0$. The two long-range orders we are interested in are the ones corresponding to a "chessboard" configuration for the $s_x$ or the $n_x$.

\begin{figure}
\begin{center}
\mbox{\psfig{figure=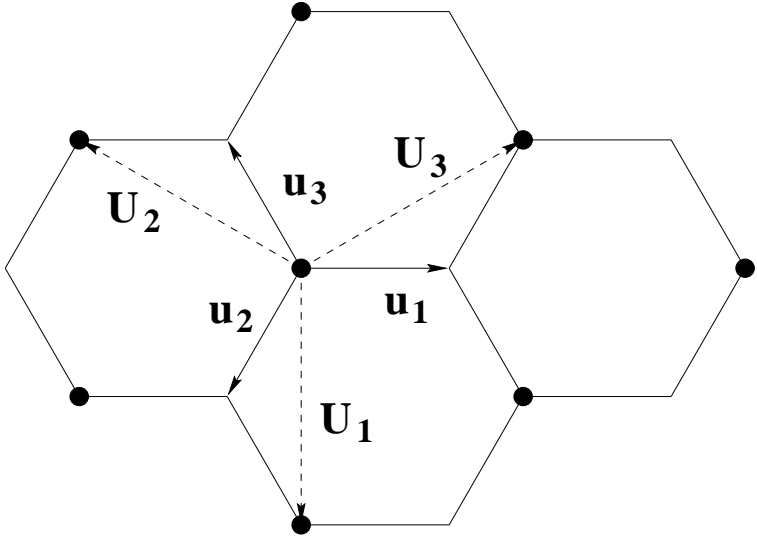,width=10cm,angle=0}}
\end{center}
\caption{The hexagonal lattice.}
\label{fig61}
\end{figure}

The hexagonal lattice is composed of two sublattices $A$ and $B$. In figure \ref{fig61}, we have represented the sites of $A$ as black dots and we have defined the vectors $u_1$, $u_2$, $u_3$ and $U_1$, $U_2$ and $U_3$. It is natural to define the analog of the Fourier transform of a local observable $O_x$ using a basis of eigenfunctions of the Hamiltonian $\sum_{x,y\in\Lambda}t_{xy}c_x^+c_y$ for a single particle hopping on the hexagonal lattice. The eigenfunctions and eigenvalues are parametrized by a wave vector $k$ in the first Brillouin zone of the triangular lattice defined by $A$ (one has to notice that this triangular lattice has $\vol/2$ sites and thus we have also $\vol/2$ $k$'s in the first Brillouin zone) and by another quantum number $\tau=\pm 1$. The eigenenergies are given by $\epsilon_{k,\tau}=\tau E_k$
with
\be
E_k=\sqrt{3+2(\cos(k\cdot U_1)+\cos(k\cdot U_2)+\cos(k\cdot U_3))},
\end{equation}
and the eigenfunctions are 
\ba
\Psi_{k,\tau}(x)&=\frac{\e{-\ii k\cdot x}}{\sqrt{\vol}},\hspace{1cm}x\in A\nonumber\\
\Psi_{k,\tau}(x)&=\frac{\tau\lambda_k\e{-\ii k\cdot x}}{\sqrt{\vol}},\hspace{1cm}x\in B
\end{align}
where
\be
\lambda_k=\sum_{j=1}^3\frac{\e{\ii k\cdot u_j}}{E_k}.
\end{equation}
They satisfy the orthogonality relation
\be
\braket{\Psi_{k,\tau}}{\Psi_{k',\tau'}}=\delta_{kk'}\delta_{\tau\tau'}.
\end{equation}
We define the Fourier transform on the hexagonal lattice as
\be
O_{k,\tau}=\sum_{x\in\Lambda}\Psi_{k,\tau}(x)O_x.
\label{hol102}
\end{equation}
The long-range orders we are interested in are the one in the two-point correlation functions $\expval{s_{-k,\tau}s_{k,\tau}}$ and $\expval{(n_{-k,\tau}-1/2)(n_{k,\tau}-1/2)}$ at $k=0$ and $\tau=-1$. The reason is that at $k=0$ and $\tau=-1$, we have 
\be
s_{0,-1}=\frac{1}{\sqrt{\vol}}\sum_{x\in A}s_x-\frac{1}{\sqrt{\vol}}\sum_{x\in B}s_x
\end{equation}
since $\lambda_0=1$. Long-range order in this last operator will indicate the presence of a chessboard configuration. 
We will prove that there exist $\epsilon_1$ and $\epsilon_2$ independent of $\Lambda$ such that
\be
\frac{1}{\vol}\expval{s_{0,-1}s_{0,-1}}=\frac{1}{\vol^2}\expval{(\sum_{x\in A}s_x-\sum_{x\in B}s_x)(\sum_{y\in A}s_y-\sum_{y\in B}s_y)}\geq \epsilon_1>0
\label{hol25}
\end{equation}
and 
\bm
\frac{1}{\vol}\expval{\rho_{0,-1}\rho_{0,-1}}=\\
\frac{1}{\vol^2}\expval{(\sum_{x\in A}(n_x-1/2)-\sum_{x\in B}(n_x-1/2))(\sum_{y\in A}(n_y-1/2)-\sum_{y\in B}(n_y-1/2))}\geq \epsilon_2>0
\label{hol26}
\end{multline}
for large $\Lambda$ and zero temperature. The global strategy is exactly the same as for the square lattice in two dimensions. We prove the coexistence of the two long-range orders, find upper bounds for the Duhamel two-point functions, then for the two-point correlation functions and finally we use the sum rule 
\be
\sum_{k}\sum_{\tau=\pm 1}\expval{\rho_{-k,\tau}\rho_{k,\tau}}=\sum_{x\in\Lambda}\expval{(n_x-1/2)^2}=\frac{\vol}{4}
\label{sumruhexa}
\end{equation}
to conclude the proof. To prove the first equality in (\ref{sumruhexa}), one uses the fact that $\lambda_k\lambda_{-k}=\lambda_k\lambda_{k}^*=1$. Let us now indicate the main steps of the proof insisting on the points that differ from the square lattice case.
 
\vspace{5mm}
\begin{center}
{\bf Coexistence of the two long-range orders}
\end{center}
\vspace{5mm}

For the coexistence of the two long-range orders and the relation between the Duhamel two-point functions, we have the commutation relation
\be
[\ham,p_x]=\ii S_x+\ii U(n_y-1/2)
\end{equation}
where 
\ba
S_x&=(\kappa+12\alpha)s_x+4\alpha\sum_{i=1}^3s_{x+u_i},\hspace{1cm}x\in A\nonumber\\
S_x&=(\kappa+12\alpha)s_x+4\alpha\sum_{i=1}^3s_{x-u_i},\hspace{1cm}x\in B.
\end{align}
In Fourier transform, we have
\be
S_{k,\tau}=[\kappa+4\alpha(3+\tau E_k)]s_{k,\tau}
\end{equation}
and for the point $k=0$, $\tau=-1$ in which we are specially interested, $S_{0,-1}=\kappa s_{0,-1}$. With these facts, it is easy to prove the coexistence of the two long-range orders (\ref{hol25}) and (\ref{hol26}). Then, by (\ref{coex1}), we obtain the relation between the Duhamel two-point functions
\bm
\left([\kappa+4\alpha(3+\tau E_k)]
\sqrt{\duha{s_{-k,\tau}}{s_{k,\tau}}}-U\sqrt{\duha{\rho_{-k,\tau}}{\rho_{k,\tau}}}\right)^{2}\\
\leq\frac{1}{\beta}[\kappa+4\alpha(3+\tau E_k)].
\label{hol27}
\end{multline}

\newpage
\vspace{5mm}
\begin{center}
{\bf Infrared bounds}
\end{center}
\vspace{5mm}

To derive an infrared bound for $\duha{S_{-k,\tau}}{S_{k,\tau}}$ we proceed as in eqs. (\ref{ineqh})-(\ref{duha-s}). For the hexagonal lattice one can bring (\ref{hol10}) into a reflection positive form when the flux through each plaquette is zero. The analog of (\ref{analog}) is 
\be
\duha{\sum_{x\in A}\sum_{i=1}^3h^*_{x,x+u_i}(s_x+s_{x+u_i})}{\sum_{y\in A}\sum_{j=1}^3h_{y,y+u_j}(s_y+s_{y+u_j})}\leq\frac{1}{4\alpha\beta}\sum_{x\in A}\sum_{i=1}^3\abs{h_{x,x+u_j}}^2.
\label{hol24}
\end{equation}
We now choose $h_{x,x+u_j}$ for $x\in A$ to be (for each $k$ in the first Brillouin zone and each $\tau$)
\be
h_{x,x+u_j}=\frac{1}{\sqrt{\vol}}\e{-\ii k\cdot x}(1+\tau\lambda_k\e{-\ii k\cdot u_j}).
\end{equation}
Inserting this form into (\ref{hol24}), we get
\be
\duha{s_{-k,\tau}}{s_{k,\tau}}\leq \frac{1}{4\alpha\beta(3+\tau E_k)}.
\label{hol28}
\end{equation}
We can transfer this information onto an inequality on the Duhamel two-point functions for the electrons using (\ref{hol27}):
\ba
g\sqrt
{\duha{\rho_{-k,\tau}}{\rho_{k,\tau}}}&\leq \frac{\kappa+4\alpha(3+\tau E_k)}{\sqrt{4\alpha\beta(3+\tau E_k)}}+
\sqrt{\frac{\kappa+4\alpha(3+\tau E_k)}{\beta}}\nonumber\\
&\leq\frac{\kappa+24\alpha}{\sqrt{\alpha\beta(3+\tau E_k)}}
\end{align}
leading to
\be
\duha{\rho_{-k,\tau}}{\rho_{k,\tau}}\leq \frac{(\kappa+24\alpha)^2}{g^2\alpha\beta(3+\tau E_k)}
\isdefby \frac{B_{k,\tau}}{\beta}.
\label{hol29}
\end{equation}

\vspace{5mm}
\begin{center}
{\bf Inequalities on two-point correlation functions}
\end{center}
\vspace{5mm}

As for the square lattice, the upper bound for the two-point correlation function in the electronic density is given by
\be
\expval{\rho_{-k,\tau}\rho_{k,\tau}}\leq
\frac{1}{2}\sqrt{B_{k,\tau}C_{k,\tau}}\coth\sqrt{\frac{\beta^2C_{k,\tau}}{4B_{k,\tau}}}
\end{equation}
where $B_{k,\tau}$ is the bound given in (\ref{hol29}) and $C_{k,\tau}$ is the upper bound on the double commutator 
\be
\expval{[n_{-k},[H,n_k]]}=\frac{1}{\vol}\sum_{x,y\in\Lambda}t_{xy}(2-\tau\e{-\ii k\cdot(x-y)}\lambda_k-\tau\e{\ii k\cdot(x-y)}\lambda_{-k})
\expval{c_x^+c_y}\leq 12\isdefby C_{k,\tau}.
\label{integral2}
\end{equation}
Finally, in the zero temperature limit $\beta\rightarrow\infty$ and for large volumes, we get
\be
\frac{1}{\vol}\expval{\rho_{0,-1}\rho_{0,-1}}\geq \frac{1}{4}-\frac{1}{(2\pi)^2}\int\dd^2 k\sum_{\tau=\pm 1}
\frac{1}{2}\sqrt{\frac{12(\kappa+24\alpha)^2}{g^2\alpha(3+\tau E_k)}}.
\end{equation}
Evaluating the integrals in (\ref{integral2}) leads to the final condition on the hexagonal lattice for zero temperature
\be
\frac{\kappa+24\alpha}{g\sqrt{\alpha}}< 0.18.
\end{equation}
The qualitative behavior in the parameters $g$, $\alpha$ and $\kappa$ is the same as for the square lattice. 

\section{Conclusion}

In the present work we have provided a rigorous proof of the stability of the Peierls instability at half-filling when quantum and thermal fluctuations are taken into account. Our results are valid for dimensions greater or equal to two for a modified Holstein model. The regime investigated here is limited to strong coupling but is independent of the strength of the quantum fluctuations. Of course, we expect that for weak coupling the mass $m$ of the oscillators should play a role but our method does not give any information about this situation. An important drawback of infrared bounds used here is that we cannot treat the one dimensional case.

One would also like to treat the standard Holstein model where the dispersion relation is flat. However we do not know how to obtain the appropriate infrared bound because it is not apparent how to couple the phonon field $s_x$ to the $h_{xy}$ variables. Another strategy would be to use so-called chessboard estimates to perform a sort of Peierls argument. We have carried out this program for a simplified version of the Holstein model where the Einstein oscillators at each site $x\in\Lambda$ are replaced by two levels systems modeled with Pauli matrices. More precisely, in the Hamiltonian, $s_x$ is replaced by $\sigma_x^{(3)}$ and $p_x^2/2m$ by $\epsilon\sigma_x^{(1)}$ ($\epsilon\sim 1/m$). For such a model, it is found that in two dimensions the Peierls instability occurs for all "electron-phonon" coupling for weak enough quantum fluctuations $\epsilon$ \cite{mp2}.

Another important issue is the effect of electron spin and electron-electron Coulomb interaction. Most of the analysis carried out here is still valid if we consider spin and add to the Hamiltonian a Hubbard term $U\sum_{x\in\Lambda}(n_{x\uparrow}-1/2)(n_{x\downarrow}-1/2)$. The coexistence theorem is still valid when applied for $s_k$ and $\rho_k=\frac{1}{\vol}\sum_{x\in\Lambda}(n_{x\uparrow}+n_{x\downarrow}-1)$. Also since the Hubbard term is reflection positive at half-filling we may deduce again infrared bounds for $\duha{s_{-k}}{s_k}$ and $\duha{\rho_{-k}}{\rho_k}$. The extra information that needs to be controlled is a lower bound for 
\be
\sum_k\expval{\rho_{-k}\rho_k}=\sum_{x\in\Lambda}\expval{(n_{x\uparrow}+n_{x\downarrow}-1)^2}=2\sum_{x\in\Lambda}\expval{n_{x\uparrow}n_{x\downarrow}}.
\label{fin}
\end{equation}
In other words, one needs to prove that there is a finite density of doubly occupied sites in order to obtain a Peierls instability. When $U\rightarrow\infty$, $\expval{n_{x\uparrow}n_{x\downarrow}}\rightarrow 0$ so that we do not expect to have a Peierls instability and therefore it should be present only for moderate Coulomb interaction. Let us notice that if (\ref{fin}) is of the order of $\gamma\vol$ for some $\gamma >0$, then the analysis presented in this work automatically implies that at half-filling there is a CDW period two state. 

\vspace{1cm}

\begin{center}
{\bf Acknowledgements}
\end{center}

We would like to thank C. Gruber for discussions. The work of C.-A. Piguet was supported by the Swiss National Science Foundation.


\end{document}